\documentclass[acus]{JAC2000}

\def\Journal#1#2#3#4{{#1} {\bf #2}, #3 (#4)}

\def\NPA{{\rm Nucl. Phys.} A}
\def\NPB{{\rm Nucl. Phys.} B}
\def\PLB{{\rm Phys. Lett.}  B}
\def\PRL{\rm Phys. Rev. Lett.}
\def\PRD{{\rm Phys. Rev.} D}
\def\PRC{{\rm Phys. Rev.} C}

\def\la{\langle}
\def\ra{\rangle}

\def\al{\alpha}

\def\be{\begin{equation}}
\def\ee{\end{equation}}
\def\bea{\begin{eqnarray}}
\def\eea{\end{eqnarray}}
\input{psfig}

\usepackage{graphicx}

\begin{document}
\title{
\centering 
NEW RROGRESS IN TIME-LIKE EXCLUSIVE PROCESSES
}

\author{Chueng-Ryong Ji, Department of Physics, North Carolina
State University, NC 27695-8202, USA\\
Ho-Meoyng Choi, Department of Physics, Carnegie Mellon
University, PA 15213, USA}

\maketitle
\begin{abstract}
We discuss a necessary nonvalence contribution in timelike exclusive
processes. Following a Schwinger-Dyson type of approach, we relate 
the nonvalence contribution to an ordinary light-front wave function
that has been extensively tested in the spacelike exclusive processes.
A complicate multi-body energy denominator is exactly cancelled in summing
the light-front time-ordered amplitudes. Applying our method to
$K_{\ell3}$ and $D^0\to K^- \ell^+ \nu_l$ where a rather substantial
nonvalence contribution is expected, we find not only an improvement
in comparing with the experimental data but also a covariance(i.e.
frame-indepndence) of existing light-front constituent quark model.
\end{abstract}

\section{INTRODUCTION}
As discussed in this PEP-N meeting, the facilities that copiously
produce the lower-lying mesons such as $K$ and $D$ can provide 
a lot of rich physics as exciting as the new and upgraded $B$-meson 
factories promise. To fulfill such excitement, however, intensive 
theoretical studies should be accompanied in the analyses of exclusive
meson decays and form factors. Thus, more and more scrutinized model 
analyses are called for. 

Perhaps, one of the most popular formulations for the analysis of
exclusive processes may be provided in the framework of light-front
(LF) quantization~\cite{BPP}.
In particular, the Drell-Yan-West ($q^+=q^0+q^3=0$)
frame has been extensively used in the calculation of various electroweak
form factors and decay processes~\cite{Ja2,CJ1,Kaon,KCJ}.
As an example, only the parton-number-conserving (valence) Fock state
contribution is needed in $q^+=0$ frame when the ``good" component
of the current,
$J^+$ or ${\bf J}_{\perp}=(J_x,J_y)$, is used for the spacelike
electromagnetic form factor calculation of pseudoscalar mesons.
The LF approach may also provide a bridge between the two
fundamentally different pictures of hadronic matter, i.e. the
constituent quark model (CQM) (or the quark parton model) closely
related to the experimental observations and the quantum chromodynamics
(QCD) based on a covariant non-abelian quantum field theory.
The crux of possible connection between the two pictures is the rational
energy-momentum dispersion relation that leads to a relatively simple
vacuum structure. There is no spontaneous creation of massive fermions
in the LF quantized vacuum. Thus, one can immediately obtain a
constituent-type picture, in which all partons in a hadronic state are
connected directly to the hadron instead of being simply disconnected
excitations (or vacuum fluctuations) in a complicated medium.
A possible realization of chiral symmetry breaking in the LF vacuum
has also been discussed in the literature~\cite{Wilson}.

\begin{figure}[htbp]
\centerline{ \psfig{figure=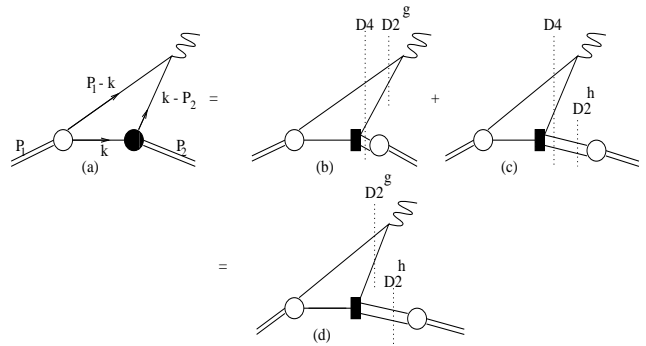,height=1.8in,width=3.3in}}
\caption{Effective treatment of the LF nonvalence amplitude.}
\label{fig1}
\end{figure}

On the other hand, the analysis of timelike exclusive processes (or
timelike $q^2>0$ region of bound-state form factors) remained
as a rather significant challenge in the LF approach. In principle, the
$q^+\neq0$ frame can be used to compute the timelike processes but
then it is inevitable to encounter the particle-number-nonconserving
Fock state (or nonvalence) contribution.
The main source of difficulty in CQM phenomenology
is the lack of information on the non-wave-function vertex(black blob
in Fig.~\ref{fig1}(a)) in the nonvalence diagram arising from the 
quark-antiquark
pair creation/annihilation. The non-wave-function vertex(black blob)
was recently also called the embedded state~\cite{BCJ}.
This should contrast with the white blob representing the usual LF
valence wave function.

In principle, there is a systematic program
laid out by Brodsky and Hwang~\cite{BH} to include the
particle-number-nonconserving
amplitude to take into account the nonvalence contributions.
However, such program requires to find all the higher Fock-state wave
functions while there has been relatively little progress in computing
the basic wave functions of hadrons from first principles.
Recently, a method of analytic continuation from the spacelike region has
also been suggested to generate necessary informations in the timelike
region without encountering a direct calculation of the nonvalence
contribution~\cite{Anal}.
Even though some explicit example has been presented
for manifestly covariant theoretical models,
this method has not yet been implemented to more realistic
phenomenological models.

In this talk, we thus present an alternative way of handling the nonvalence
contribution. Our aim of new treatment~\cite{JC} is to make the program more
suitable for the CQM phenomenology specific to the low momentum transfer
processes. Incidentally, the light-to-light ($K_{\ell3}$) and heavy-to-light
($D^0\to K^-\ell^+\nu_\ell$) decays involving rather low momentum transfers
bear a substantial contribution from the nonvalence part and their
experimental data are better known than other semileptonic processes with
large momentum transfers.
Including the nonvalence contribution, our results on
$K_{\ell3}$ and $D^0\to K^-\ell^+\nu_\ell$ not only show a definite
improvement in comparison with experimental data but also exhibit a
covariance (i.e frame-independence) of our approach.

This talk is organized as follows. In Section 2, we present the 
non-wave-function vertex in the nonvalence diagram in terms of
light-front vertex functions, utilizing the covariant 
Bethe-Salpeter(BS) model of (3$+$1)-dimensional fermion field theory.
The nonvalence part of the weak form factors for $0^-\to 0^-$
semileptonic decays is expressed in terms
of light-front vertex functions of a hadron and a gauge boson.
The link operator connecting $(n-1)$-body to $(n+1)$-body in a
Fock state representation is obtained by an analytic continuation
of the usual BS amplitude. 
We also show that the complicated $(n+2)$-body energy denominators
are exactly cancelled in summing the 
light-front time-ordered diagrams.
In Section 3, we show our numerical results for $K_{\ell3}$ and 
$D^0\to K^-\ell^+\nu_\ell$ decays. Conclusions follow in Section 4.
\section{New effective treatment}

\subsection{$0^-\to0^-$ Semileptonic Decays}

The semileptonic decay of $Q_{1}\bar{q}$ bound state with four-momentum
$P^\mu_1$ and mass $M_1$ into another $Q_{2}\bar{q}$ bound state with
$P^\mu_2$ and $M_2$ is governed by the weak current, viz.,
\bea{\label{eq:K1}}
J^{\mu}(0)&=&\la P_{2}|\bar{Q_{2}}\gamma^{\mu}Q_{1}|P_{1}\ra\nonumber\\
&=& f_{+}(q^{2})(P_{1}+P_{2})^{\mu} + f_{-}(q^{2})q^{\mu},
\eea
where $q^\mu=(P_{1}-P_{2})^\mu$ is the four-momentum transfer to
the lepton pair ($\ell\nu$) and $m^{2}_\ell\leq q^2\leq (M_{1}-M_{2})^{2}$.
The covariant three-point Bethe-Salpeter (BS) amplitude of the total current
$J^\mu(0)$ in Eq.~(\ref{eq:K1}) may be given by 
\bea{\label{eq:Jmu}}
J^{\mu}(0)&=&iN_c\int \frac{d^{4}k}{(2\pi)^4}
\frac{H^{\rm cov}_1H^{\rm cov}_2S^\mu}{(p^{2}_{1}-m^{2}_{1}+i\epsilon)
(p^{2}_{2}-m^{2}_{2}+i\epsilon)}\nonumber\\
&&\;\;\;\;\;\;\times\frac{1}{(p^{2}_{\bar{q}}-m^{2}_{\bar{q}}+i\epsilon)},
\eea
where $N_c$ is the color factor, $H^{\rm cov}_{1[2]}$ is
the covariant intitial[final] state meson-quark vertex function
that satisfies the BS equation, and
$S^\mu = {\rm Tr}[\gamma_{5}({\not\! p}_{1}+m_{1})\gamma^\mu
({\not\! p}_{2}+m_{2})\gamma_{5} (-{\not\!p}_{\bar{q}}+m_{\bar{q}})]$.
The quark momentum variables are given by $p_1=P_1 - k$, $p_2=P_2-k$, and
$p_{\bar{q}}=k$.

As shown in the literature~\cite{BCJ}, the LF energy integration reveals
an explicit correspondence between the sum of LF time-ordered
amplitudes and the original covariant amplitude.
For instance, performing the $k^-$ pole integration,
we obtain the LF currents, $J^\mu_V$ and $J^\mu_{NV}$ corresponding to
the usual LF valence diagram and the nonvalence diagram shown in
Fig.~\ref{fig1}(a), respectively.
Since $H^{\rm cov}_{2}$ satisfies the BS equation, we iterate
$H^{\rm cov}_{2}$ once and perform its LF energy integration to find the
corresponding LF time-ordered diagrams 
Figs.~\ref{fig1}(b) and~\ref{fig1}(c) after the iteration.
The similar idea of iteration in a Schwinger-Dyson (SD) type of approach
was presented in Ref.~\cite{BJS} to pin down the LF bound-state equation
starting from the covariant BS equation.

\begin{figure}[htbp]
\centerline{\psfig{figure=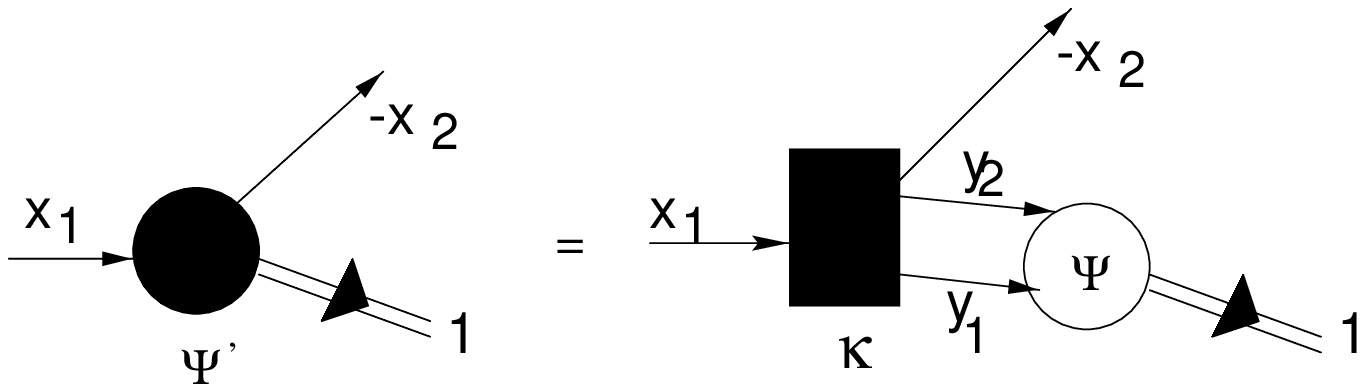,height=1.0in,width=3.2in}}
\caption{Non-wave-function vertex(black blob) linked to an ordinary
LF wave function(white blob).}\label{fig2}
\end{figure}

Comparing the LF time-ordered expansions before and after the
iteration, we realize that the following link between the
non-wave-function vertex (black blob) and the
ordinary LF wave function (white blob) as shown in Fig.~\ref{fig2}
naturally arises, i.e.,
\bea\label{eq:SD}
&&(M^2-{\cal M}^{2}_0)\Psi'(x_i,{\bf k}_{\perp i})
\nonumber\\
&&=\int[dy][d^2{\bf l}_{\perp}]
{\cal K}(x_i,{\bf k}_{\perp i};y_j,{\bf l}_{\perp j})
\Psi(y_j,{\bf l}_{\perp j}),
\eea
where $M$ is the mass of outgoing meson and ${\cal M}^{2}_0=(m^2_1+{\bf
k}^2_{\perp 1})/x_1 - (m^2_2+{\bf k}^2_{\perp 2})/(-x_2)$ with
$x_1 = 1-x_2 > 1$ due to the kinematics of the non-wave-function vertex.

We note that Eq.~(\ref{eq:SD}) essentially takes the same form as
the LF bound-state equation (similar to the LF projection of BS
equation) except the difference in kinematics(e.g. $-x_2 >0$ for the
non-wave-function vertex).
Incidentally, Einhorn~\cite{Einhorn} also discussed the extension of the
LF BS amplitude in 1+1 QCD to a non-wave-function vertex similar to
what we obtained in this work.

In the above procedure, we also find that the
four-body energy denominator $D_4$ is exactly cancelled in the sum of LF
time-ordered amplitudes as shown in Figs.~\ref{fig1}(b) 
and~\ref{fig1}(c), i.e., $1/D_4D^g_2 + 1/D_4D^h_2 =1/D^g_2D^h_2$.
We thus obtain the amplitude identical to the nonvalence
contribution in terms of ordinary LF wave functions of gauge boson($W$)
and hadron (white blob) as drawn in Fig.~\ref{fig1}(d). This method, however,
requires to have some relevant operator depicted as the black
square(${\cal K}$) in Fig.~\ref{fig2}(See also Fig.~\ref{fig1}(d)), 
that is in general dependent on the involved momenta connecting 
one-body to three-body sector.
We now present some details of kinematics in the semileptonic decay
processes to discuss a reasoning of how we handle the nonvalence
contribution involving the momentum-dependent ${\cal K}$
for relatively small momentum transfer processes such
as $\pi_{e3}$, $K_{\ell3}$ and $D\to K\ell\nu$.

\subsection{Kinematics and Model Description}
Our calculation is performed in purely longitudinal momentum
frame~\cite{JC,Zero} 
where $q^+>0$ and ${\bf P}_{1\perp}={\bf P}_{2\perp}=0$
so that the momentum transfer square $q^2=q^+q^->0$ is time-like. 
One can then easily obtain $q^2$ in terms of the momentum fraction
$\alpha=P^{+}_{2}/P^{+}_{1}=1-q^{+}/P^{+}_{1}$
as $q^{2}=(1-\alpha)(M^{2}_{1}-M^{2}_{2}/\al)$.
Accordingly, the two solutions for $\alpha$ are given by
\be\label{apm}
\alpha_{\pm}=\frac{M_{2}}{M_{1}}\biggl[
\frac{ M^{2}_{1}+M^{2}_{2}-q^{2}}{2M_{1}M_{2}}
\pm \sqrt{\biggl(\frac{ M^{2}_{1}+M^{2}_{2}-q^{2}}
{2M_{1}M_{2}}\biggr)^{2}-1} \biggr].
\ee
The $+(-)$ sign in Eq.~(\ref{apm}) corresponds to the daughter meson
recoiling in the positive(negative) $z$-direction relative to
the parent meson. At zero recoil($q^{2}=q^{2}_{\rm max}$) and
maximum recoil($q^{2}=0$), $\alpha_{\pm}$ are given by
\begin{eqnarray}\label{alimit}
&&\alpha_{+}(q^{2}_{\rm max})=
\alpha_{-}(q^{2}_{\rm max})=\frac{M_{2}}{M_{1}},
\nonumber\\
&&\alpha_{+}(0)=1,\hspace{0.5cm}
\alpha_{-}(0)=\biggl(\frac{M_{2}}{M_{1}}\biggr)^{2}.
\end{eqnarray}
In order to obtain the form factors $f_{\pm}(q^{2})$ which are
independent of $\alpha_{\pm}$, defining
$J^+(0)|_{\alpha=\alpha_{\pm}} \equiv 2P_{1}^{+}H^{+}(\alpha_{\pm})$
from Eq.~(\ref{eq:K1}), we obtain
\be\label{fpm}
f_{\pm}(q^{2})=\pm \frac{(1\mp \alpha_{-})H^{+}(\alpha_{+}) -
(1\mp \alpha_{+})H^{+}(\alpha_{-})}{\alpha_{+}-\alpha_{-}}.
\ee
The form factors $f_{+}(q^2)$ and $f_{-}(q^2)$ are related to 
the scalar form factor $f_0(q^2)$ in  the following way:
\be\label{f0}
f_{+}(0)=f_{0}(0),\hspace{.2cm}f_{0}(q^{2})= f_{+}(q^{2})
+ \frac{q^{2}}{M^{2}_1-M^{2}_2}f_{-}(q^{2}).
\ee
The differential decay rate for $0^-\to 0^-$ semileptonic decay
is given by~\cite{Kaon}
\begin{eqnarray}\label{GF}
\frac{d\Gamma}{dq^{2}}\hspace{-.3cm}&=&\hspace{-.3cm} 
\frac{G^{2}_{F}}{24\pi^{3}}|V_{q_1\bar{q}_2}|^{2}
K_{f}(q^{2})(1-\frac{m^{2}_{l}}{q^{2}})^{2}\nonumber\\
\hspace{-0.3cm}&\times&\hspace{-0.3cm}
\biggl\{
[K_{f}(q^{2})]^{2}(1+\frac{m^{2}_{l}}{2q^{2}})|f_{+}(q^{2})|^{2}
\nonumber\\
\hspace{-0.3cm}&+&\hspace{-.3cm}
M^2_1(1-\frac{M^2_2}{M^2_1})^{2}\frac{3}{8}
\frac{m^{2}_{l}}{q^{2}}|f_{0}(q^{2})|^{2}\biggr\},
\end{eqnarray}
where $G_{F}$ is the Fermi constant, $V_{q_1\bar{q}_2}$ is
the element of the Cabbibo-Kobayashi-Maskawa(CKM) mixing matrix and
the factor $K_{f}(q^{2})$ is given by
\be\label{Kf}
K_{f}(q^{2})=\frac{1}{2M_1}
\biggl[ (M^2_1+M^2_2-q^{2})^{2}-4M^{2}_1M^{2}_2
\biggr]^{1/2}.
\ee

With the iteration procedure Eq.~(\ref{eq:SD}) in this $q^+>0$ frame,
the results for the ``$+$"-component of the current
$J^\mu$ in Eq.~(\ref{eq:Jmu}) are given by
\be{\label{eq:JV}}
J^{+}_{\rm V}=\frac{N_c}{16\pi^3}
\int^{\alpha}_{0}dx\int d^{2}{\bf k}_{\perp}
\frac{\Psi_{i}(x,{\bf k}_{\perp}) S^{+}_{V}
\Psi_{f}(x',{\bf k}_{\perp})}{x(1-x)(1-x')},
\ee
and
\bea{\label{eq:JNV}}
J^{+}_{NV}\hspace{-0.3cm}&=&\hspace{-0.3cm}\frac{N_c}{16\pi^3}
\int^{1}_{\alpha}dx
\int d^{2}{\bf k}_{\perp}\frac{\Psi_i(x,{\bf k}_{\perp})S^{+}_{NV}}
{x(1-x)(x'-1)}\Psi_g(x,{\bf k}_{\perp})\nonumber\\
\hspace{-0.3cm}&\times&\hspace{-0.3cm}
\int \frac{1}{y(1-y)}\int d^2{\bf l}_{\perp}
{\cal K}(x,{\bf k}_{\perp};y,{\bf l}_{\perp})
\Psi_{f}(y,{\bf l}_{\perp}),
\eea
where Eq.~(\ref{eq:SD}) has been used for the nonvalence wave
function at the black blob as shown in 
Fig.~\ref{fig2}(see also Fig.~\ref{fig1}(d)).

The ordinary LF vertex functions(white blob in Fig.~\ref{fig1})
in Eqs.~(\ref{eq:JV}) and~(\ref{eq:JNV}) are given by 
\bea\label{wf12}
\Psi_{i}=
\frac{h^{\rm LF}_{1}}{M^{2}_{1}-M^{2}_{01}},\;
M^{2}_{01}=\frac{m^{2}_1 + {\bf k}^2_{\perp}}{1-x}
+ \frac{m^{2}_{\bar{q}} + {\bf k}^2_{\perp}}{x},\nonumber\\
\Psi_{f}=
\frac{h^{\rm LF}_{2}}{M^{2}_{2}-M^{2}_{02}},\;
M^{2}_{02}=\frac{m^{2}_2 + {\bf k}^2_{\perp}}{1-x'}
+ \frac{m^{2}_{\bar{q}} + {\bf k}^2_{\perp}}{x'},
\eea
where $x=k^+/P^{+}_{1}$, $x'=x/\alpha$. 
The $\Psi_g$ in Eq.~(\ref{eq:JNV}) corresponds to the light-front
energy denominator(i.e. $D_g$ in Fig.~\ref{fig1}(d)) and its
explicit form is given by
\be\label{gaugeWF}
\Psi_g(x,{\bf k}_\perp)=
\frac{1}{\alpha\biggl[\frac{q^2}{1-\alpha} -
\biggl(\frac{{\bf k}^2_\perp + m^2_1}{1-x}
+\frac{{\bf k}^2_\perp + m^2_2}{x -\alpha}\biggr)
\biggr]}.
\ee
We call $\Psi_g$ the light-front vertex function of a gauge 
boson~\footnote{ While one can in principle also consider the BS
amplitude for $\Psi_g$, we note that such extension does not alter our
results within our approximation in this work because both hadron and gauge
boson should share the same kernel.}.

In Eqs.~(\ref{eq:JV}) and~(\ref{eq:JNV}), 
the trace terms 
$S^+_{V}(p^{-}_{\bar{q}}=k^{-}_{\rm on})
=(4P^{+}_{1}/x')\{ {\bf k}^{2}_{\perp}
+ [x m_1  + (1-x)m_{\bar{q}}][x'm_2 + (1-x')m_{\bar{q}}]\}$
and $S^{+}_{NV}(p^-_1=p^-_{1\rm on})= S^{+}_{V}(p^-_i=p^-_{i{\rm on}})
+ 4p^{+}_{\rm 1on}p^{+}_{\rm 2on}(p^{-}_{\bar{q}}
- p^{-}_{\bar{q}{\rm on}})$
correspond to the product of initial and
final LF spin-orbit wave functions that are uniquely determined by a
generalized off-energy-shell Melosh transformation. Here,
the subscript (on) means on-mass-shell and the instantaneous part of
nonvalence diagram corresponds to
$4p^{+}_{\rm 1on}p^{+}_{\rm 2on}(p^{-}_{\bar{q}}
- p^{-}_{\bar{q}{\rm on}})$ in $S^+_{NV}$.
While the LF vertex function $h^{LF}_{1[2]}$ formally stems from
$H^{\rm cov}_{1[2]}$, practical informations on the radial wave function
$\Psi_{i[f]}(x,{\bf k}_{\perp})$ (consequently $h^{LF}_{1[2]}$) can be
obtained by LF CQM. The details of our variational procedure to determine
both mass spectra and wave functions of pseudoscalar mesons were recently
documented in Refs.~\cite{CJ1,Kaon} along with an extensive test of the model
in the spacelike exclusive processes. 
The same model is used in this work, i.e.,
comparing the LF vertex functions $\Psi$ in Eq.~(\ref{wf12})
with our light-front wave function given by Ref.~\cite{CJ1,Kaon}, 
we identify
\be\label{LFvertex}
\Psi(x,{\bf k}_\perp)=\biggl(\frac{8\pi^3}{N_c}\biggr)^{1/2}
\biggl(\frac{\partial k_z}{\partial x}\biggr)^{1/2}
\frac{[x(1-x)]^{1/2}}{\tilde{M}_0}\phi(x,{\bf k}_\perp),
\ee
where $\tilde{M}_0=\sqrt{M^2_0-(m_q-m_{\bar q})^2}$ and
the Jacobian of the variable tranformation
${\bf k}=(k_z,{\bf k}_\perp)\to
(x,{\bf k}_\perp)$ is obtained as $\partial k_z/\partial x=M_0/[4x(1-x)]$
and the radial wave function is given by
\be\label{radial}
\phi({\bf k}^2)=\biggl(\frac{1}{\pi^{3/2}\beta^3}\biggr)^{1/2}
\exp(-{\bf k}^2/2\beta^2),
\ee
which is normalized as $\int d^3k|\phi({\bf k}^2)|^2=1$.
Substituting Eqs.~(\ref{LFvertex}) and~(\ref{radial}) into 
Eqs.~(\ref{eq:JV}) and~(\ref{eq:JNV}), one can obtain the valence
and nonvalence contributions to the weak form factors for 
$0^-\to 0^-$ semileptonic decays in light-front quark model.

While the relevant operator ${\cal K}$ is in general
dependent on all internal momenta ($x,{\bf k}_{\perp},y,{\bf l}_{\perp}$),
a sort of average on ${\cal K}$ over $y$ and ${\bf l}_{\perp}$
in Eq.~(\ref{eq:JNV}) which we define as
$G_{P_1P_2}\equiv\int[dy][d^2{\bf l}_{\perp}]
{\cal K}(x,{\bf k}_\perp;y,{\bf l}_\perp)\Psi_f(y,{\bf l}_{\perp})$
is dependent only on $x$ and ${\bf k}_{\perp}$.
Now, the range of the momentum fraction $x$ depends on the
external momenta for the embedded states.
As shown in Eq.~(\ref{eq:JNV}), the lower bound of $x$
for the kernel in the nonvalence contribution is given by $\alpha$ which
has the value $\alpha=M_2/M_1$ at the maximum $q^2$.
As the mass difference between the primary and secondary mesons gets
smaller, not only the range of $q^2$ is reduced but also $\alpha$ gets
closer to 1. Perhaps, the best experimental process for such limit
may be the pion beta decay $\pi^\pm \to \pi^0 e^\pm {\bar \nu_e}$, where
our numerical prediction $f_-(0)/f_+(0)= -3.2\times 10^{-3}$
following the treatment presented in this work is in an excellent
agreement with $-3.5\times 10^{-3}$ obtained by the
method proposed by Jaus~\cite{Ja1} including the 
zero-modes~\cite{Chang,BH,Zero}.
In Ademollo-Gatto's SU(3) limit~\cite{AG}, the $q^2$ range of
the nonvalence contribution shrinks to zero and $\alpha$ becomes
precisely 1.
However, even if $\alpha$ is not so close to 1, the initial wavefunction
$\Psi_i(x,{\bf k}_{\perp})$ plays the
role of a weighting factor in the nonvalence contribution and enfeeble
the contribution from the region of $x$ near 1. Thus,
for the processes that we discuss in this talk,
the effective $x$ region for the nonvalence contribution
is quite narrow. Similarly, the region of the
transverse momentum ${\bf k}_{\perp}$ is also limited only up to the
scale of hadron size due
to the same weighting factor $\Psi_i(x,{\bf k}_{\perp})$. Here,
we thus approximate $G_{P_1P_2}$ as a constant and examine the validity
of this approximation by checking the frame independence of our
numerical results.

For the check of frame-independence,
we also compute the ``$+$" component of the current $J^\mu_{D}$
in the Drell-Yan-West ($q^+=0$) frame where only valence contribution
exists. 
Since the form factor $f_+(q^2)$ obtained
from $J^+_D$ in $q^+=0$ frame is immune to the zero-mode
contribution~\cite{Kaon,BH,Zero,Ja1},
the comparison of $f_+(q^2)$ in the two completely different frames
(i.e. $q^+=0$ and $q^+\neq0$) would reveal the validity of existing model
with respect to a covariance.
The comparison of $f_-(q^2)$, however, cannot give a meaningful test
of covariance because of the zero-mode complication as noted in
Ref.~\cite{Ja1}. Indeed, the difference between the two ($q^+=0$ and
$q^+\neq0$) results of $f_-(q^2)$ amounts to the zero-mode
contribution.

\section{Numerical Results}

\begin{figure}[htbp]
\centerline{\psfig{figure=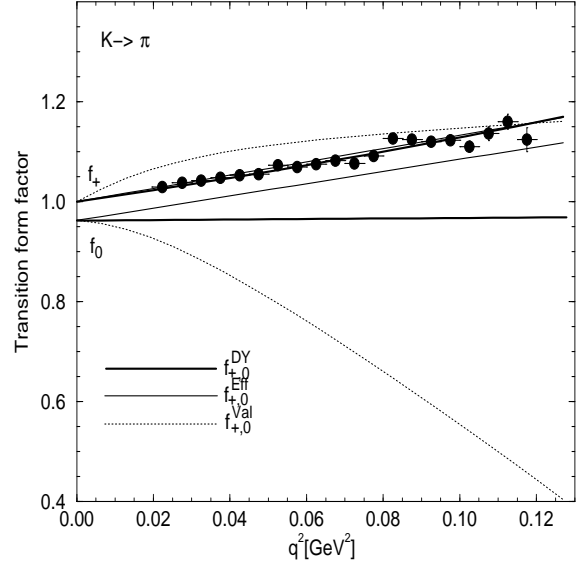,height=3.5in,width=3.5in}}
\caption{The weak form factors for
$K^{0}_{\ell3}$ compared with the experimental data~\protect\cite{Apo}.}
\label{Kl3}
\end{figure}

In our numerical calculation for the processes of $K_{\ell3}$ and
$D^0\to K^-\ell^+\nu$ decays, we use the linear potential parameters presented
in Ref.~\cite{Kaon}. 
In Fig.~\ref{Kl3}, we show the weak form factors $f_+(q^2)$ and
$f_0(q^2)$ for $K^{0}_{\ell3}$ decays.
The thick solid lines are our analytic solutions obtained from
the $q^+=0$ frame; note here again that the lower thick solid line
($f_0$) in Fig.~\ref{Kl3} is only the partial result without including the
zero-mode contribution while the upper thick solid line
($f_+$ immune to the zero-mode) is the full result.
The thin solid lines are the full results of our effective calculations
with a constant ($G_{K\pi}$=3.95) fixed by the normalization of $f_+$ at
$q^2=0$ limit.
For comparison, we also show only the valence
contributions(dotted lines) in $q^+\neq0$ frame.
As expected, a clearly distinguishable nonvalence contribution is found.
Following the popular linear parametrization~\cite{Data}, we
plot the results of our effective solutions(thin solid lines) using
$f_{i}(q^2)=f_{i}(q^2=m^{2}_{\ell})(1+\lambda_{i}q^2/M^{2}_{\pi^+})(i=+,0)$.
In comparison with the data, the same normalization as the data
$f_{+}(0)=1$~\cite{Apo} was used in Fig.~\ref{Kl3}.
Our effective solution(upper thin solid line) is not only
in a good agreement with the data~\cite{Apo}
but also almost identical to that in $q^+=0$ frame(upper thick
solid line) indicating the frame-independence of our model.
Note also that the difference in $f_0(q^2)$ between $q^+\neq0$(lower
thin solid line) and $q^+=0$(lower thick solid line) frames amounts
to the zero-mode contribution.

\begin{table*}[htbp]
\centering
\caption{ Model predictions for the parameters of $K^{0}_{\ell3}$ decays.
The decay width is in units of $10^{6}$ s$^{-1}$. The
used CKM matrix is $|V_{us}|=0.2196\pm0.0023$~\protect\cite{Data}.}
\begin{tabular}{|c|c|c|c|}
\hline
&Effective & $q^+=0$ & Experiment\\
\hline
$f_{+}(0)$ & 0.962 [0.962] & 0.962 [0.962] & \\
\hline
$\lambda_{+}$& 0.026 [0.083] & 0.026 [0.026] & $0.0288\pm0.0015
[K^{0}_{e3}]$\\
\hline
$\lambda_{0}$& 0.025 [$-0.017$]& 0.001 [$-0.009$]
& $0.025\pm0.006[K^{0}_{\mu3}]$\\
\hline
$\xi_{A}$& $-0.013$ [$-1.10$]& $-0.29 [-0.41]$
& $-0.11\pm0.09[K^{0}_{\mu3}]$\\
\hline
$\Gamma(K^{0}_{e3})$ & $7.3\pm0.15$
&  $7.3\pm0.15$ & 7.5$\pm$0.08\\
\hline
$\Gamma(K^{0}_{\mu3})$ & $4.92\pm0.10$
&  $4.66\pm0.10$ & 5.25$\pm$0.07\\
\hline
\end{tabular}
\end{table*}
In comparison with experimental data, we summarized our results of
several experimental observables in Table~1;
i.e. the actual value of $f_+(0)$, the slopes $\lambda_+$
[$\lambda_0$] of $f_+(q^2)$ [$f_0(q^2)$] at $q^2=0$,
$\xi_{A}$=$f_-(0)/f_+(0)$, and the decay rates $\Gamma(K^{0}_{e3})$
and $\Gamma(K^{0}_{\mu3})$.
In the second column of Table~1,
our full results including nonvalence contributions are presented along
with the valence contributions in the square brackets. In the third
column of Table~1, the results in $q^+=0$ frame are presented with[without]
the instantaneous part. As one can see in Table 1, adding the nonvalence
contributions clearly improves the results of $\lambda_0$, i.e.
our full result of $\lambda_0$= 0.025 is in an excellent agreement with
the data, $\lambda^{\rm Exp.}_0$= 0.025$\pm$0.006.
Since the lepton mass is small except in the case of the $\tau$
lepton, one may safely neglect the lepton mass in the decay rate
calculation of the heavy-to-heavy and heavy-to-light transitions.
However, as one can see from the improved result for $K_{\mu3}$ 
decay rate, the reliable calculation of $f_0(q^2)$ is required 
especially for $K_{\mu 3}$ since the muon($\mu$) mass is not negligible, 
even though the contribution of $f_0(q^2)$ is negligible for
$K_{e3}$ case. 

\begin{figure}
\centerline{\psfig{file=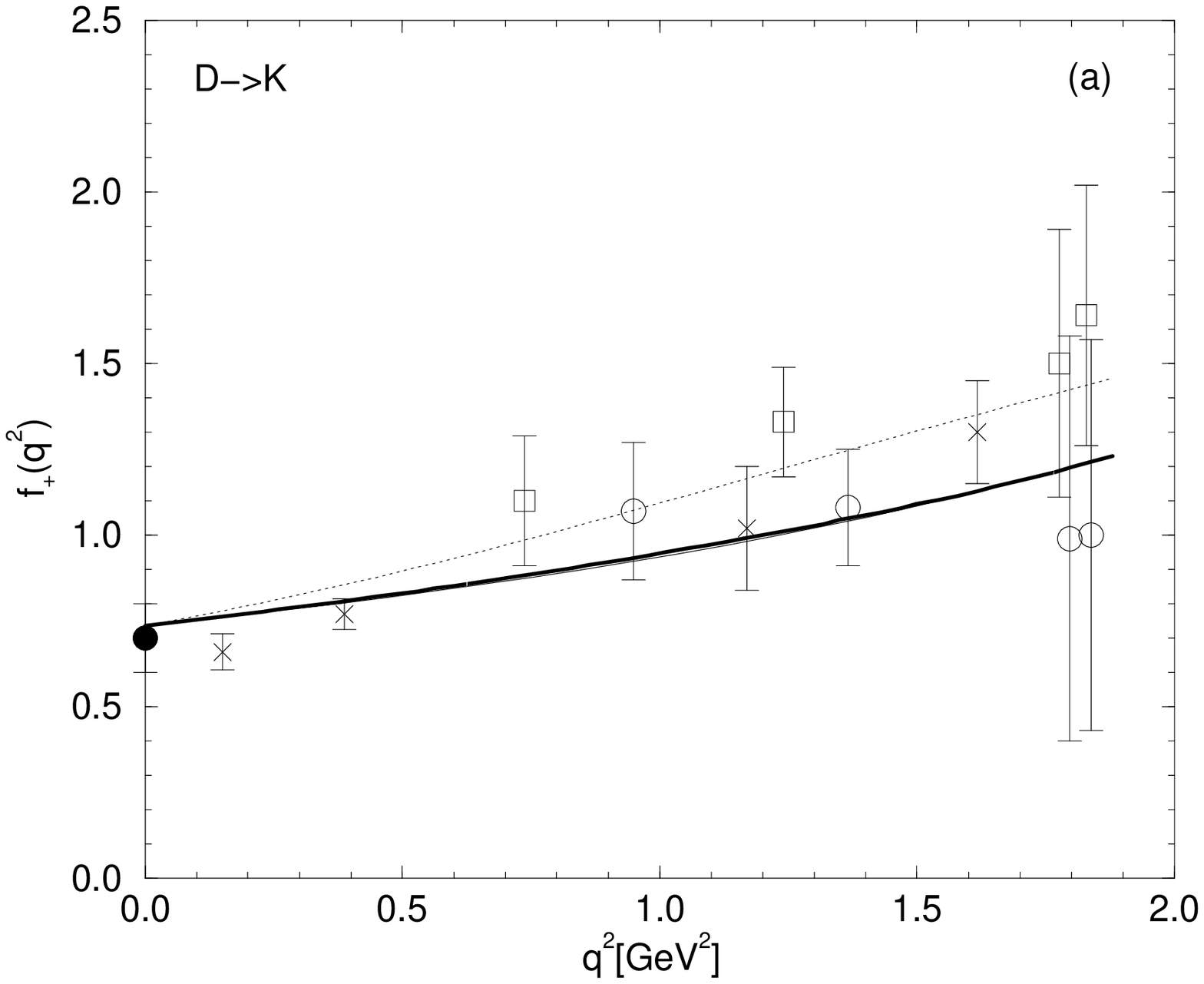,height=3.5in,width=3.5in}}
\centerline{\psfig{file=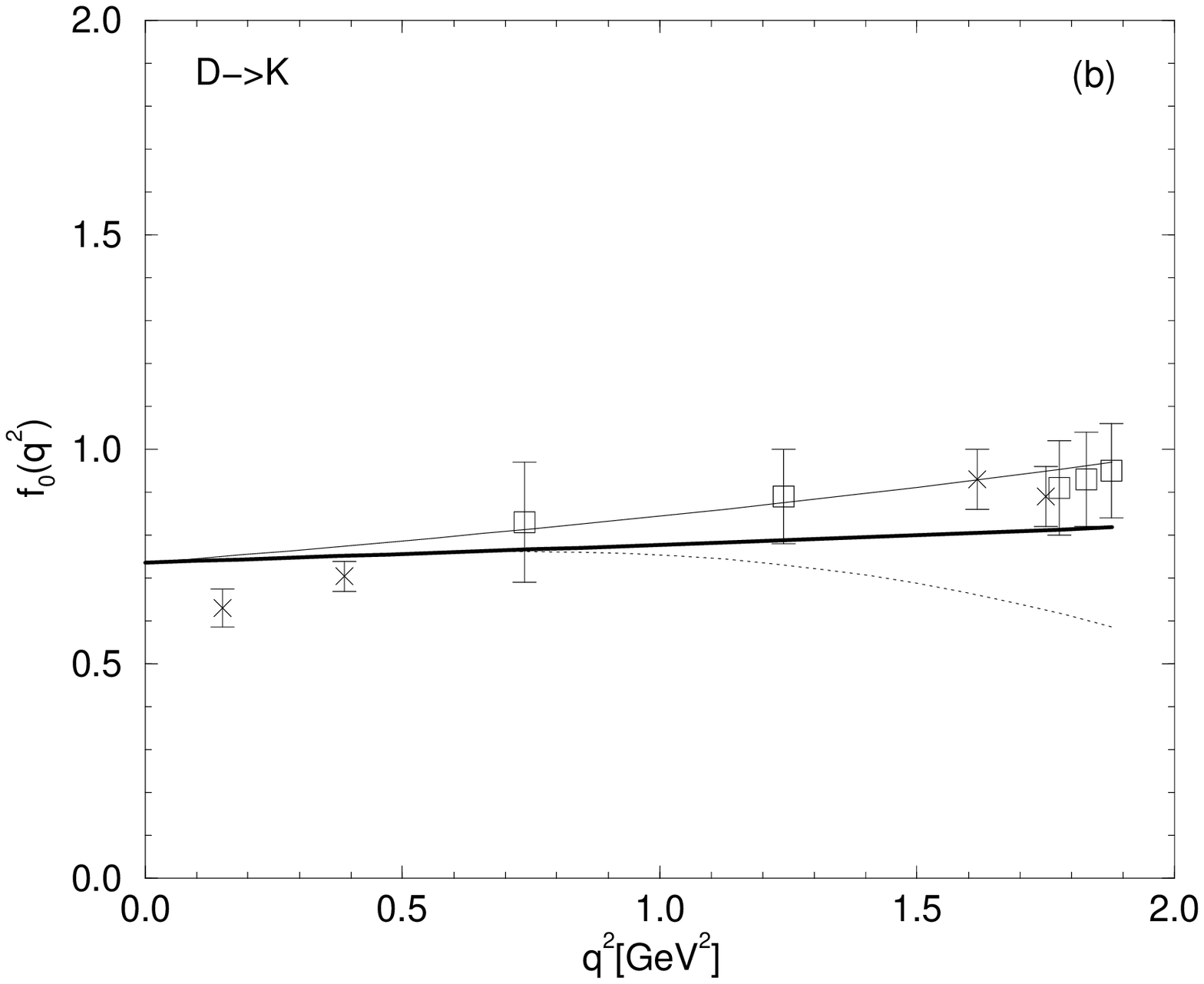,height=3.5in,width=3.5in}}
\caption{The weak form factors for $D\to K$
transition. The same line codes are used as in Fig.~3.}
\end{figure}

In Figs.~4(a,b), we show the weak form factors for $D^0\to K^-\ell^+\nu$
decays and compare with the experimental data~\cite{Data}(full dot) with an
error bar at $Q^2=0$ as well as the lattice QCD results~\cite{Ber}(circle
and square)
and~\cite{Bow}(cross). All the line assignments are same as in Fig.~3.
In Fig.~4(a), the thin solid line of our full result in $q^+\neq0$ is not
visible because it exactly coincides with the thick solid line of
the result in $q^+ = 0$ confirming the frame-independence of our
calculations. Our value of $f_+(0)$=0.736 is also within the error bar of
the data~\cite{Data}, $f^{\rm Exp.}_+(0)$=0.7$\pm$0.1. In Fig.~4(b),
the difference between the thin and thick solid lines is the measure
of the zero-mode contribution to $f_0(q^2)$ in $q^+=0$ frame.
The form factors obtained from our effective
calculations($G_{DK}$=3.5) are also plotted with the usual
parametrization of pole dominance model, i.e.
$f_{+(0)}(q^2)=f_{+(0)}(0)/(1-q^2/M^2_{1^{-}(0^+)})$.
Our pole masses turn out to be $M_{1^-}$=2.16 GeV and $M_{0^+}$=2.79 GeV,
respectively, and we note that $M_{1^-}$=2.16 GeV is in a good
agreement with the mass of $D^*_{s}$,i.e. 2.1 GeV.
Using CKM matrix element $|V_{cs}|=1.04\pm 0.16$~\cite{Data},
our branching ratios ${\rm Br}(D^{0}_{e3}) = 3.73\pm 1.24$ and
${\rm Br}(D^{0}_{\mu3}) = 3.60\pm 1.19$ are also comparable with the
experimental data $3.64\pm 0.18$ and $3.22\pm 0.17$~\cite{Data}, 
respectively. 

\begin{figure}
\centerline{\psfig{file=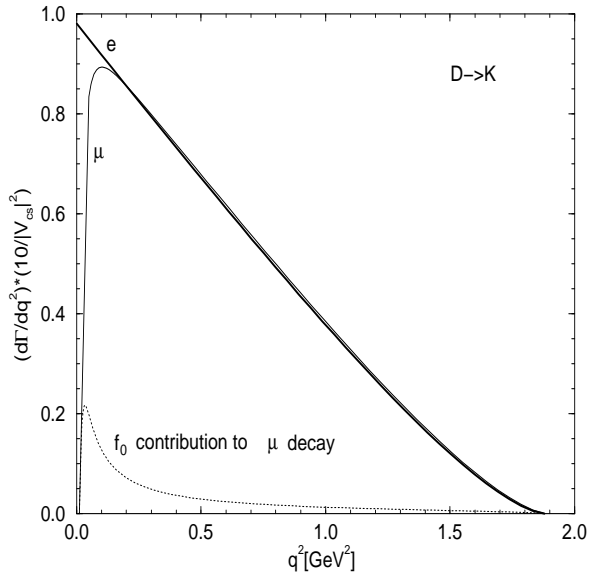,height=3.5in,width=3.5in}}
\caption{The differential decay rates for $D^0\to K^-\ell^+\nu_\ell$
transition.}\label{Drate}
\end{figure}
In Fig.~\ref{Drate}, we show the differential decay rates for
$D^0\to K^-e^+\nu_e$ and $D^0\to K^-\mu^+\nu_\mu$ transitions obtained
from our effective solutions. 
As in the case of $K_{\ell3}$ decays, we were able to 
evaluate the $f_0(q^2)$ contribution to the total decay
rate for $D^0\to K^-\mu^+\nu_\mu$ process in a more reliable
manner although its contribution is more suppressed than
the $K_{\mu3}$ case.  

\section{Conclusion}
In summary, we presented an effective treatment of the LF
nonvalence contributions crucial in the timelike exclusive processes.
Using a SD-type approach and summing the LF time-ordered amplitudes,
we obtained the nonvalence contributions in terms of ordinary LF
wavefunctions of gauge boson and hadron
that have been extensively tested in the spacelike exclusive processes.
Including the nonvalence contribution, our results show a definite
improvement in comparison with experimental data on $K_{\ell3}$
and $D^0\to K^-\ell^+\nu_\ell$ decays. Our result on $\pi_{e3}$
is also consistent with the result obtained by other method.
Furthermore, the frame-independence of our
results indicate that a constant $G_{P_1 P_2}$ is an approximation
appropriate to the small momentum transfer processes. 
A similar conclusion was drawn in a
recent application of our method to the skewed quark distributions
of the pion at small momentum transfer region~\cite{CJK}. 
Applications to the heavy-to-light decay processes involving large
momentum transfers would require an improvement on
this approximation perhaps guided by the perturbative QCD approach.
Consideration along this line is underway.

\begin{center}
{\bf Acknowledgements}
\end{center}
The work of CRJ was supported in part by the US DOE under 
contracts DE-FG02-96ER40947 and that of HMC by the NSF grant 
PHY-00070888.
The North Carolina Supercomputing Center and the National Energy Research
Scientific Computer Center are also acknowledged for the grant of
supercomputer time.

\end{document}